\documentstyle[prl,aps,amssymb,epsfig,graphicx,tabularx,multicol]{revtex}

\newfont{\ensmathquatorze}{msbm10 scaled 1400}
\newfont{\ensmathonze}{msbm10 scaled 1100}
\newfont{\ensmathdix}{msbm10}
\newfont{\ensmathneuf}{msbm10 scaled 833}
\newfont{\ensmathhuit}{msbm10 scaled 694}
\newfam\ensmathfam                        
\textfont\ensmathfam=\ensmathonze        
\scriptfont\ensmathfam=\ensmathdix       
\scriptscriptfont\ensmathfam=\ensmathhuit

\def\be{\begin{equation}}
\def\ee{\end{equation}}

\def\bea{\begin{eqnarray}}
\def\eea{\end{eqnarray}}
\def\beann{\begin{eqnarray*}}
\def\eeann{\end{eqnarray*}}

\newcommand{\ket}[1]{|\kern.3ex#1\kern.3ex\rangle}
\newcommand{\bra}[1]{\langle\kern.3ex #1 \kern.3ex|}

\newcommand{\smean}[1]{\langle #1 \rangle} 

\newcommand{\EXP}[1]{{\mbox{\large e}}^{#1}}         
\renewcommand{\sinh}[1]{\mathop{\mathrm{sh}}\nolimits #1} 

\def\I{{\rm i}}                  
\def\D{{\rm d}}                  
\def\Dc{{\rm D}}                 

\newcommand{\drond}[2]{\frac{\partial #1}{\partial #2}} 

\newcommand\ab{{\alpha\beta}}

\newcommand{\diagram}[3]{\raisebox{#3}{\includegraphics[scale=#2]{#1}}}

\begin{document}

\title{Weak localization in multiterminal networks of diffusive wires}

\author{Christophe Texier$^{(a,b)}$ and Gilles Montambaux$^{(b)}$}
 
\date{December 2, 2003} 

\maketitle	

\medskip

{\small
$^{(a)}$Laboratoire de Physique Th\'eorique et Mod\`eles Statistiques.
Universit\'e Paris-Sud,
B\^at. 100, F-91405 Orsay Cedex, France.

$^{(b)}$Laboratoire de Physique des Solides.
Universit\'e Paris-Sud,
B\^at. 510, F-91405 Orsay Cedex, France.
}

\medskip

\begin{abstract}
We study the quantum transport through networks of diffusive wires 
connected to reservoirs in the Landauer-B\"uttiker formalism. 
The elements of the conductance matrix are computed by the diagrammatic 
method. We recover the combination of classical resistances and obtain the 
weak localization corrections.
For arbitrary networks, we show how the cooperon must be properly weighted 
over the different wires. Its nonlocality is clearly analyzed.
We predict a new geometrical effect that may change the sign of the weak 
localization correction in multiterminal geometries.
\end{abstract}

\medskip

\noindent
PACS~: 73.23.-b~; 73.20.Fz~; 72.15.Rn





\begin{multicols}{2}

How to increase transmission with weak localization~? This is one the 
question addressed in this letter devoted to a general description of 
transport in networks of quasi-one-dimensional weakly disordered wires.
At the classical level, the network is equivalent to a network of classical
resistances, which gives the dominant contribution to the conductances.
Additionally, there exists a small correction, the ``weak localization'' 
correction, originating from quantum interferences.
For the conductivity, it reads \cite{GorLarKhm79,AltKhmLarLee80}~:
$\Delta\sigma(\vec r)=-\frac{e^2}{\pi}P_c(\vec r,\vec r)$ 
(with $\hbar=1$, without spin factor),
where  the cooperon $P_c$ is the contribution from pairs of time reversed
trajectories to the return probability.
Networks are particularly well suited to study interference effects
and the first experimental studies of weak localization in such systems,
performed on honeycomb lattices \cite{PanChaRamGan84}, 
showed the oscillations predicted by Altshuler-Aronov-Spivak (AAS) 
\cite{AltAroSpi81}. 
These experimental results were well fitted by the theory of Dou{\c c}ot \& 
Rammal (DR) \cite{DouRam85} whose starting point is a uniform integration of 
$\Delta\sigma(\vec r)$ over the wires of the network \cite{footnote}.
However, although this procedure is meaningful for a translation invariant
system, it is not valid in general for networks. In this letter we  
demonstrate that the correct expression of the weak localization correction
to the resistance ${\cal R}$ of a network of wires  $(\mu\nu)$ of 
lengths $l_{\mu\nu}$ and section $s$ is instead~:
\be\label{heuristres}
\Delta{\cal R}
= \sum_{(\mu\nu)} \drond{{\cal R}^{\rm cl}}{R_{\mu\nu}^{\rm cl}} 
\Delta R_{\mu\nu} 
\mbox{ with }
\frac{\Delta R_{\mu\nu}}{R^{\rm cl}_{\mu\nu}}=
\int_{(\mu\nu)}\hspace{-0.5cm}{\D\vec r}\,
\frac{-\Delta\sigma(\vec r)}{s\,l_{\mu\nu}\,\sigma_0}
\ee
where ${\cal R}^{\rm cl}$ is the classical resistance obtained from the
classical laws of combination of resistances $R^{\rm cl}_{\mu\nu}$.
The sum runs over all wires $(\mu\nu)$. $\sigma_0$ is the Drude conductivity.
This result has a simple structure~: it could be obtained from small variations
$\Delta R_{\mu\nu}$ in the classical expression ${\cal R}^{\rm cl}$ of the
resistance.
However it is highly non trivial since, due to
nonlocality, it is not possible to get a quantum formula for the resistance
of the network as a function of quantum resistances of the wires. It is in
fact not even possible to define a quantum resistance for a wire,
independently of the whole network.
Eq.~(\ref{heuristres}) shows that a uniform integration of the cooperon
is applicable only to  regular networks in which the weights of the wires
$\partial{\cal R}^{\rm cl}/\partial R_{\mu\nu}^{\rm cl}$ are all equal, 
like in the experiments of Ref. \cite{PanChaRamGan84}. 
We show that in a multiterminal geometry, some of these weights can change 
in sign.
This can lead to a {\it change in sign of the weak localization correction.}

\begin{figure}
\begin{center}
\includegraphics[scale=0.9]{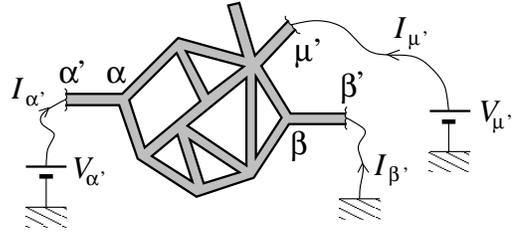}
\end{center}
\caption{A network of diffusive wires. The network is connected at 
         the vertices $\alpha'$, $\beta'$ and $\mu'$ to external reservoirs 
         through which some current is injected in the network. 
         \label{fig:fig1}}
\end{figure}

As on figure \ref{fig:fig1}, we consider networks that can be connected 
to external contacts (here the contacts $\alpha'$, $\beta'$ and $\mu'$).
The transmission probability between two contacts (the conductance matrix
element, up to a factor $e^2/h$), when averaged over the disorder, can be 
written as
$\smean{T_{\alpha'\beta'}}=T^{\rm cl}_{\alpha'\beta'}
+\Delta T_{\alpha'\beta'}+\cdots$, where
the first term is the classical result (Drude conductance) and the second  
is the weak localization correction. Our aim is to give a systematic way to
compute the weak localization contribution in terms of matrices encoding 
the information on the network (topology, lengths of the wires, magnetic 
fluxes).
The most natural approach to describe transport in networks is the 
Landauer-B\"uttiker approach.
However some difficulties related to the question of current conservation 
are more conveniently overcome in the Kubo formulation. 
We first recall some features of the transport theory of weakly disordered 
metals in the Kubo approach and eventually use the connection to the 
Landauer-B\"uttiker formalism that we apply to the case of  networks.
Finally we consider several examples. 

\medskip

\noindent{\it Classical transport}.
The classical transport is described by two contributions~: 
a Drude contribution (short range) and the contribution from the 
diffuson (ladder diagrams) which is long range.
Then, in the diffusion approximation, we have \cite{KanSerLee88}~:
\bea\label{classcon}
&\langle\sigma_{ij}(\vec r,\vec r\,')\rangle_{\rm class}
= \diagram{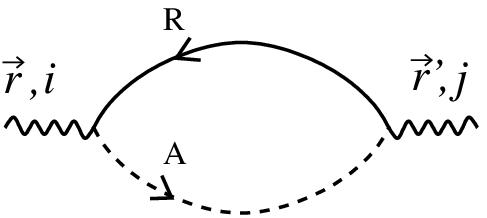}{0.5}{-0.4cm} 
+ \diagram{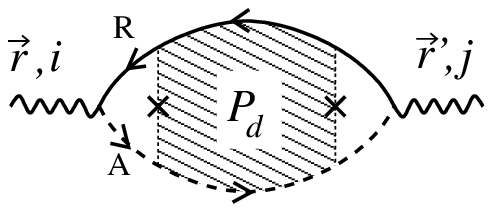}{0.5}{-0.6cm}\nonumber\\
&=\sigma_{\rm 0}\,
\left[
  \delta_{ij}\:{\delta}(\vec r-\vec r\,')
 -\nabla_i\nabla'\!\!_j P_d(\vec r,\vec r\,')
\right] = \sigma_{\rm 0}\: \phi_{ij}(\vec r,\vec r\,')
\:,\eea
where the diffuson $P_d$ is solution of the equation
$-\Delta P_d(\vec r,\vec r\,')=\delta(\vec r-\vec r\,')$.
An important requirement of a transport theory is to satisfy current
conservation $\nabla_i\sigma_{ij}(\vec r,\vec r\,')=0$, what the classical
conductivity (\ref{classcon}) does.
The question of boundary conditions is not a trivial one. In most works,
it is argued, by analogy with the problem of classical diffusion, that 
$\vec{n}\cdot\vec\nabla P_d|_{\partial {\cal D}}=0$ on the reflecting 
boundaries of the domain ${\cal D}$ ($\vec{n}$ is the vector normal to 
the surface), while $P_d|_{\partial {\cal D}}=0$
on the boundaries where the disordered system is connected to a reservoir.
This latter is described by a region free of disorder.
However it was shown in \cite{Ish78,HasStoBar94,AkkMon04} that the diffuson 
vanishes at a distance of order $\ell_e$ inside the reservoir.
$\ell_e$ is the elastic mean free path.
In the case of a quasi-1d wire, the diffuson vanishes at a distance
$x_d=\alpha_d\ell_e/2$, where  $d$ is the dimension with
$\alpha_1=2$, $\alpha_2=\pi/2$, $\alpha_3=4/3$.

\begin{figure}
\begin{center}
\includegraphics[scale=0.7]{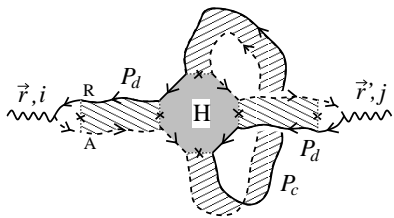}
\end{center}
\caption{A long range diagram for the weak localization.\label{fig:sigwl}}
\end{figure}

\noindent{\it Current conserving weak localization correction}.
The weak localization correction involves the 
cooperon which, in a magnetic field $\vec{\cal B}=\vec\nabla\times\vec A$, 
is solution of~:
\be\label{eqdifcoop}
\left[\frac{1}{L_\varphi^2}-\left(\vec\nabla-2\I e\vec A\right)^2\right]
P_c(\vec r,\vec r\,') 
= \delta(\vec r-\vec r\,')
\:.\ee
The contribution of the cooperon reads \cite{GorLarKhm79,AltKhmLarLee80}~:
\bea\label{GLK}
&\smean{\sigma_{ij}(\vec r,\vec r\,')}_{\rm cooperon}
 = \diagram{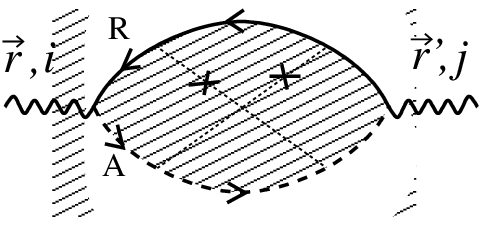}{0.5}{-0.6cm} \nonumber\\
&\hspace{1cm}
= - \frac{e^2}{\pi}\,\delta_{ij}\,\delta(\vec r-\vec r\,')\,
P_c(\vec r,\vec r)
\:.\eea
It is clear that the additional contribution of the cooperon 
(\ref{GLK}) does not respect current conservation. 
In the same way that the classical conductivity is built of short range 
(Drude) and long range (diffuson) contributions, the weak localization 
correction contains long range terms additionally to the short range 
contribution (\ref{GLK}). A possible way to build a long range diagram is to
dress the current vertices with diffusons, like on figure \ref{fig:sigwl}.

The connection of the cooperon and the diffusons is realized by introducing a
Hikami box \cite{Hik81}. Dressing either the left current line with a 
diffuson, or the right one or both, we obtain three diagrams that add to 
the contribution of the cooperon.
However these dominant diagrams are not the only ones needed to satisfy 
current conservation \cite{HasStoBar94}. 
This is related to the fact that the diagram \ref{fig:sigwl}, which 
seems at first sight to be the only long range diagram needed to compute 
the weak localization, gives a volumic divergence $V\delta(0)$ to the 
conductance (for details, see \cite{TexMonAkk04}).
This problem was first mentioned in the study of
conductivity fluctuations \cite{KanSerLee88} where a more simple procedure
to construct a current conserving theory was proposed, that avoids to
consider all the set of diagrams of \cite{HasStoBar94}.
Kane, Serota and Lee showed, for the conductivity fluctuations, that the 
expression of the current conserving conductivity is obtained by a
``convolution'' of the short range object with the function
$\phi_{ij}$ involved in (\ref{classcon}). Finally the expression of the 
weak localization correction to the nonlocal conductivity reads~:
\bea\label{KSL}
&&\smean{\Delta\sigma_{ij}(\vec r,\vec r\,')}
= \int\D\vec\rho\,\D\vec\rho\,'\,
\nonumber\\
&&\hspace{1.5cm}
\phi_{ii'}(\vec r,\vec\rho)\,\phi_{jj'}(\vec r\,',\vec\rho\,')\,
\smean{\sigma_{i'j'}(\vec \rho,\vec \rho\,')}_{\rm cooperon}
\:,\eea
which obviously respects current conservation.

\medskip

\noindent{\it Networks}.
We now consider specifically the case of networks such as the one on figure 
\ref{fig:fig1}. The transmission $T_{\alpha'\beta'}$ between the two contacts
(conductance in unit $e^2/h$)
is related to the nonlocal conductivity $\sigma_{ij}(\vec r,\vec r\,')$
with $\vec r$ being integrated through the section of the contact $\alpha'$
and $\vec r\,'$ through the section of the contact $\beta'$ \cite{BarSto89}.
For quasi-1d wires, we obtain the expressions
\be\label{dfg}
T^{\rm cl}_{\alpha'\beta'}=\frac{\alpha_d N_c}{\ell_e}
P_d(\underline{\alpha}',\underline{\beta}')
\ee
where $N_c$ is the number of channels in the wires, and
\be\label{wlg}
\Delta{T}_{\alpha'\beta'}=\frac2{\ell_e^2}
\int_{\rm Network}\hspace{-0.25cm}\D x\, 
\frac{\D}{\D x}P_d(\underline{\alpha}',x)\:
P_c(x,x)\:\frac{\D}{\D x}P_d(x,\underline{\beta}')
\:,\ee
that involve the one-dimensional diffuson and cooperon, solutions of the 
diffusion equation
$\left[\gamma - \Dc_x^2 \right] P(x,x') = \delta(x-x')$,
where $\Dc_x=\D_x-2\I eA(x)$ is the covariant derivative and 
$\gamma=1/L_\varphi^2$ (for $P_d$ we set $\gamma=0$ and $A(x)=0$).
The notation $P_d(\underline{\alpha}',x)$ means that the diffuson is 
taken at a distance $\ell_e$ of the vertex $\alpha'$.
Eq.~(\ref{dfg},\ref{wlg}) assume Dirichlet boundary conditions at the 
vertices connected to reservoirs. 
Indeed, it can be shown that the correct boundary conditions mentioned 
above can be replaced by $P_d|_{\partial {\cal D}}=0$, providing to insert 
a multiplicative factor $4$, independent of $d$, as it has been done
in (\ref{dfg},\ref{wlg}). This simplifies slightly the calculations.

As a check, let us consider the case of a quasi-1d wire
of length $L$ connected to reservoirs at both sides.
We recover the Drude dimensionless conductance
$g_{\rm D}=\alpha_d N_c\ell_e/L$, and the correction
$\Delta g = -\frac{L_\varphi}{L}
\big(\coth\frac{L}{L_\varphi}-\frac{L_\varphi}{L}\big)$ which 
interpolates between $\Delta g=-1/3$ in the fully coherent limit 
($L\ll L_\varphi$), and $\Delta g=-L_\varphi/L$ for $L\gg L_\varphi$.

We introduce the adjacency matrix $a_{\alpha\beta}$ that encodes
the information on the topology of the network~:
$a_{\alpha\beta}=1$ if the vertices $\alpha$ and $\beta$ are connected by 
a wire $(\alpha\beta)$, whose length is denoted by $l_{\alpha\beta}$;
$a_{\alpha\beta}=0$ otherwise.
To construct the solution of the diffusion equation, we need to specify 
boundary conditions at the vertices. We impose continuity of $P$ and 
$\sum_\beta a_{\alpha\beta}P'_{(\ab)}(\alpha) = \lambda_\alpha P(\alpha)$~:
the adjacency matrix constrains the sum to run over the neighbouring
vertices of $\alpha$.
The parameters $\lambda_\alpha$ describe how the network is connected.
$\lambda_\alpha=0$ for an internal vertex and $\lambda_\alpha=\infty$
at the vertices connected to external reservoirs, which imposes a Dirichlet 
boundary condition. The solution for $P_c$ involves the matrix 
\cite{DouRam85,PasMon99,AkkComDesMonTex00,TexMonAkk04}~:
\bea
{\cal M}_\ab = &&
\delta_\ab
\left(\lambda_\alpha + \sqrt\gamma\sum_\mu a_{\alpha\mu}
      \coth(\sqrt{\gamma}l_{\alpha\mu})\right)
\nonumber\\ &&\hspace{1cm}
-a_\ab\frac{\sqrt\gamma\:\EXP{-\I\theta_\ab}}{\sinh(\sqrt{\gamma}l_\ab)}
\:.\eea
The diffuson is expressed in terms of the same matrix with $\gamma=0$ and 
no magnetic flux $\theta_\ab=0$~:
\be
\left({\cal M}_0\right)_\ab = 
\delta_\ab
\left(\lambda_\alpha + \sum_\mu a_{\alpha\mu}\frac1{l_{\alpha\mu}}\right)
-a_\ab\frac{1}{l_\ab}
\:.\ee
This matrix encodes the information about the classical conductances 
$\alpha_d N_c\ell_e/l_{\mu\nu}$ of each wire $(\mu\nu)$.
Note that $\lambda_{\alpha'}=\infty$ (at a reservoir)
implies that $({\cal M}^{-1})_{\mu\alpha'}=0$, $\forall$ $\mu$.
The classical conductance is given by
\be\label{RES1}
T^{\rm cl}_{\alpha'\beta'} =
\frac{\alpha_d N_c\ell_e}{l_{\alpha\alpha'}l_{\beta\beta'}} 
\left({\cal M}_0^{-1}\right)_\ab
\:.\ee
This result is only valid for $\alpha'\neq\beta'$
\cite{HasStoBar94}. It coincides with the one obtained for 
a network of classical resistances, as it should. 

In Eq.~(\ref{wlg}), we separate the integral over the network into 
contributions of the different wires~:
$\int_{\rm Network}\to\sum_{(\mu\nu)}\int_{(\mu\nu)}$.
Since the diffuson $P_d(\underline{\alpha}',x)$ is linear in $x$ on the 
wire $(\mu\nu)$, the derivatives produce coefficients depending on $(\mu\nu)$~:
\bea\label{pres2}
&&\frac{\D}{\D x}P_d(\underline{\alpha}',x\in(\mu\nu))
= \frac{\ell_e}{l_{\alpha\alpha'}l_{\mu\nu}}
\nonumber\\
&&\times\left[ 
- \left({\cal M}_0^{-1}\right)_{\alpha\mu}\,
+ \left({\cal M}_0^{-1}\right)_{\alpha\nu}
+ \delta_{\mu\alpha}\delta_{\nu\alpha'}\, l_{\alpha\alpha'}
\right]
\:.\eea
Replacing (\ref{pres2}) in (\ref{wlg}) shows explicitly the non trivial 
weights that should be attributed to each wire when integrating the cooperon 
over the wires.
These weights have a clear physical meaning~: they can be related to 
derivatives of the corresponding classical conductance with respect to 
the lengths of the wires.
From (\ref{wlg},\ref{RES1},\ref{pres2}) we obtain~:
\be\label{RES3}
\Delta T_{\alpha'\beta'} =
\frac{2}{\alpha_d N_c\ell_e} \sum_{(\mu\nu)} 
\frac{\partial T^{\rm cl}_{\alpha'\beta'}}{\partial\,l_{\mu\nu}}
\int_{(\mu\nu)}\D x\, P_c(x,x)
\:.\ee
We have demonstrated (\ref{heuristres}), since 
$
\sigma_0                          
=\frac{e^2\alpha_d N_c\ell_e}{2\pi\,s}
$.
The integral of the cooperon over the bond $(\mu\nu)$ is a {\it nonlocal} 
quantity that carries information on the whole structure of the network 
through the matrix ${\cal M}^{-1}$~:

\end{multicols}

\bea\label{RES2bis}
\int_{(\mu\nu)}\D x\,P_c(x,x)
=\frac1{2\sqrt\gamma}\bigg\{
\left[ 
\left({\cal M}^{-1}\right)_{\mu\mu}+
\left({\cal M}^{-1}\right)_{\nu\nu}
\right]
\left(
  \coth\sqrt\gamma l_{\mu\nu} 
- \frac{\sqrt\gamma l_{\mu\nu}}{\sinh^2\sqrt\gamma l_{\mu\nu}}
\right)\nonumber\\[0.25cm]
+
\left[ 
 \left({\cal M}^{-1}\right)_{\mu\nu}\EXP{\I\theta_{\mu\nu}}
+\left({\cal M}^{-1}\right)_{\nu\mu}\EXP{\I\theta_{\nu\mu}}
+\frac{\sinh\sqrt\gamma l_{\mu\nu}}{\sqrt\gamma}
\right]
\frac{-1+  \sqrt\gamma l_{\mu\nu} \coth\sqrt\gamma l_{\mu\nu} }
     {\sinh\sqrt\gamma l_{\mu\nu}} 
\bigg\}
\:.\eea

\begin{multicols}{2}

\noindent
Eqs. (\ref{RES3},\ref{RES2bis}) give the weak localization correction for 
any network. We now turn to few special cases.

\noindent{\it Incoherent networks}.
If all wires are longer than the phase coherence length 
($l_{\mu\nu}\gg L_\varphi$, $\forall$ $(\mu\nu)$), the weak localization 
correction (\ref{RES3}) involves the same length as the classical 
conductance (\ref{RES1}). If we write 
$T^{\rm cl}_{\alpha'\beta'} = {\alpha_d N_c\ell_e}/L_{\rm eff}$, then
\be
\Delta{T}_{\alpha'\beta'} \simeq 
-\frac{L_\varphi}{l_{\alpha\alpha'}l_{\beta\beta'}}
\left({\cal M}_0^{-1}\right)_{\alpha\beta} = -\frac{L_\varphi}{L_{\rm eff}}
\:.\ee
The ratio $\Delta{T}_{\alpha'\beta'}/T^{\rm cl}_{\alpha'\beta'}$ is network
independent in this case. This simple result strongly relies on the 
hypothesis of wires with equal sections \cite{TexMonAkk04}.

\begin{figure}
\begin{center}
\includegraphics[scale=0.9]{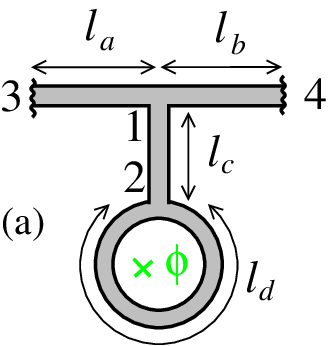}
\hspace{0.25cm}
\includegraphics[scale=0.7]{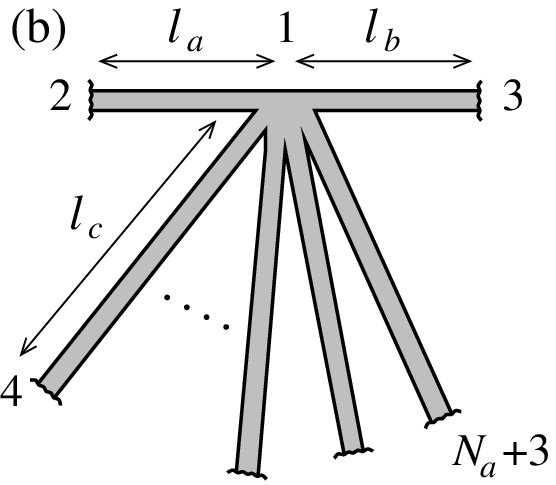}
\end{center}
\caption{Two examples of mesoscopic devices. The wavy lines indicate 
         connection to external reservoirs.\label{fig:example}}
\end{figure}

\begin{figure}
\begin{center}
\includegraphics[scale=0.5]{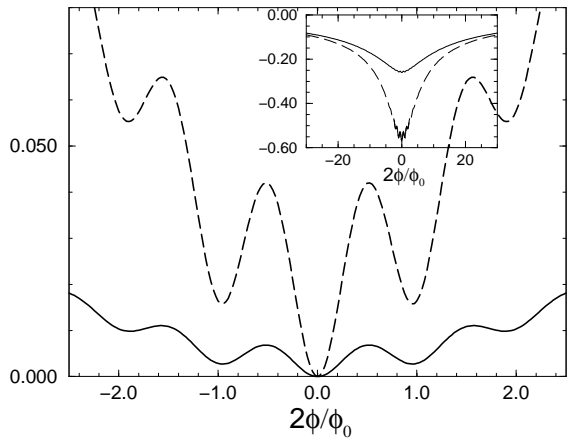}
\end{center}
\caption{Dashed line~: $\Delta\sigma/\sigma_0$ for a uniform
         integration of $P_c$.
         Continuous line~: $\Delta T_{34}/T^{\rm cl}_{34}$ 
         given by (\ref{RES3}).
         The curves have been shifted so that they coincide at $\phi=0$.
         Inset~:
         Same curves (without shift) for a higher window of 
         flux $\phi$.
         The parameters are 
         $l_a=l_b=1\:\mu$m, $l_c=0.05\:\mu$m and $l_d=5\:\mu$m.
         $W=0.19\:\mu$m, $L_\varphi(\phi=0)=1.7\:\mu$m. 
         \label{fig:ringnonlocal}}
\end{figure}

\noindent{\it A ring}.
The transport through a ring has been studied in \cite{San89}.
Here we rather consider the device on figure \ref{fig:example}.a
to illustrate the nonlocality of the weak localization.
The classical conductance (\ref{RES1}) is
$T^{\rm cl}_{34} = \frac{\alpha_d N_c\ell_e}{l_a+l_b}$. It is independent
of the presence of the arm and the ring. 
Then their weights are 0 and this part of the network 
does not contribute to the weak localization correction (\ref{RES3}). 
However, since the cooperon $P_c(x,x)$ is {\it nonlocal}, it feels the 
presence of the loop, even for $x$ in the wires $(3\leftrightarrow1)$ and 
$(1\leftrightarrow4)$.
Note that the naive uniform integration of the cooperon over the network
(DRPM) strongly overestimates the amplitude of the AAS oscillations 
(figure \ref{fig:ringnonlocal}) \cite{TexMonAkk04}.
The decrease of the weak localization at high field (inset) is due to the 
contribution of the flux to the effective phase coherence length 
$L_\varphi(\phi)$ \cite{AltAro81}.

\noindent{\it Multiterminal geometry}.
The origin of the negative weak localization $\Delta T_{\alpha'\beta'}$
lies in the negative weights, however for a multiterminal geometry some 
weights can be {\it positive}, like the weight(s)
$\partial{T^{\rm cl}_{\alpha'\beta'}}/\partial{l_{\mu\mu'}}$ of the wire(s)
connected to other terminal(s) ({\it e.g.} $\mu'$ on figure \ref{fig:fig1}).
As a first example, we consider a wire on which is plugged one long arm of
length $l_c$ connected to a third reservoir (figure \ref{fig:example}.b
for $N_a=1$).
We focus ourselves on the fully coherent limit $L_\varphi=\infty$.
The classical (Drude) conductance of this 3-terminal network is~:
$T^{\rm cl}_{23}=\frac{\alpha_d N_c\ell_e\,l_c}{l_al_b+l_bl_c+l_cl_a}$.
Then ${\partial T^{\rm cl}_{23}}/{\partial l_c}>0$.
The wire [$(2\leftrightarrow1)+(1\leftrightarrow3)$] gives a negative 
contribution to the weak localization correction whereas the arm 
$(1\leftrightarrow4)$ gives a {\it positive}
one. Introducing $l_{a/\!/b/\!/c}^{-1}=l_a^{-1}+l_b^{-1}+l_c^{-1}$, we find
$$
\Delta T_{23}
=\frac13
\left(
  -1 + \frac{l_{a/\!/b/\!/c}}{l_c} + \frac{l_{a/\!/b/\!/c}^2}{l_al_b} 
\right)
\simeq\frac13\left(-1+\frac{l_{a/\!/b}}{l_a+l_b}\right)
$$
in the limit $l_{a/\!/b}\ll l_c$.
We now consider the case of $N_a$ long arms plugged in the middle of the wire
($l_a=l_b$)  like on figure \ref{fig:example}.b , to maximize their effect
\cite{TexMonAkk04}. We obtain~:
\be\label{powl}
\Delta T_{23}\simeq\frac13\left(-1 + \frac{N_a}{4}\right)
\:,\ee 
a result valid for $l_a\ll l_c\ll L_\varphi$. 
We can now obtain a positive weak localization correction for $N_a>4$. 
This effect is purely geometrical.
Note that in the limit $l_c\gg L_\varphi$ the positive 
contribution vanishes.

\vspace{0.1cm}

\noindent{\it Conclusion.}
We have provided a general theory for the quantum transport of networks
of diffusive wires connected to reservoirs. We obtained the 
classical conductances and the weak localization corrections. 
We emphasized the importance of the weights to give to the wires
when integrating the cooperon over the network. This can lead to new
geometrical effects like a positive correction (\ref{powl}),
the physical reason being that coherent backscattering in a multiterminal
geometry can enhance a transmission.


\begin{thebibliography}{10}
\bibitem{footnote}
  The theory of DR \cite{DouRam85} was made more efficient in \cite{PasMon99} 
  (PM) by finding a simple way to express the integral of the cooperon over 
  the network  (see also \cite{AkkComDesMonTex00}). 
  PM considered thermodynamic properties for which it is correct to integrate 
  the cooperon uniformly over the network.
\end{thebibliography}

\end{multicols}

\end{document}